\documentstyle [12pt,twoside]{article}
\input{epsf.tex}
\begin{document}
\small
\par\noindent to appear in: {\it Annals of the New York Academy of Sciences}
\vskip .1in
\par\noindent
{\large\bf Phase Space Transport in Noisy Hamiltonian Systems}
\vskip .2in
\renewcommand{\thefootnote}{\alph{footnote}}
\par\noindent{HENRY E. KANDRUP\footnote{\small
HEK was supported in part by National Science Foundation Grant No.
PHY92-03333 and by Los Alamos National Laboratory through the Institute of
Geophysics and Planetary Physics. Some of the numerical calculations described 
here were facilitated by computer time provided by {\it IBM} through the 
Northeast Regional Data Center (Florida).}}
\vskip .1in
\par\noindent
{\it Department of Astronomy and Department of Physics}
\vskip .02in
\par\noindent
{\it and Institute for Fundamental Theory}
\vskip .02in
\par\noindent
{\it University of Florida, Gainesville, Florida 32611}
\par\noindent
\vskip .15in
\noindent{\small
ABSTRACT. This paper analyses the effect of 
low amplitude friction and noise in accelerating 
phase space transport in time-independent Hamiltonian systems 
that exhibit global stochasticity. Numerical experiments reveal that even 
very weak non-Hamiltonian perturbations can dramatically increase the rate 
at which an ensemble of orbits penetrates obstructions like cantori or 
Arnold webs, thus accelerating the approach towards an invariant measure, 
i.e., a microcanonical population of the accessible phase space region. 
An investigation of first passage times through cantori leads to three 
conclusions, namely: (i) that, at least for white noise, the detailed form 
of the perturbation is unimportant, (ii) that the presence or absence of 
friction is largely irrelevant, and (iii) that, overall, the  amplitude of 
the response to weak noise scales logarithmically in the amplitude of the 
noise.}
\vskip .15in
\centerline{\bf WHY CONSIDER FRICTION AND NOISE?}
\vskip .15in
In general, very weak non-Hamiltonian perturbations will only have very weak 
effects on properties of flows in time-independent Hamiltonian systems which
are integrable or near-integrable and admit no global stochasticity. This
also seems to be true for systems completely dominated by chaos where regular 
orbits are virtually nonexistent. However, low amplitude non-Hamiltonian 
perturbations {\it can} have important qualitative effects on more complex 
Hamiltonian systems that admit significant measures of both regular and 
chaotic orbits. For example, weak noise will serve as a source of extrinsic 
diffusion that can dramatically accelerate phase space transport through 
cantori (for $D=2$) or along Arnold webs (for $D{\;}{\ge}{\;}3$).

Consider, e.g., flows in two-dimensional systems. Here one knows that, in
the absence of friction and noise, cantori [1,2], fractured {\it KAM} tori
associated with the breakdown of integrability that contain a cantor set of
holes, can partition a single connected chaotic phase space region into 
separate parts which, albeit not completely disjoint, are distinct in the 
sense that a chaotic orbit starting in one part of the phase space will remain 
stuck in that part for a long time before wending its way through 
one or more holes in the cantori to access another region [3,4].
However, introducing even very weak friction and noise can dramatically 
increase the rate at which orbits pass through these holes, thus allowing 
orbits to probe the entire accessible phase space much more quickly. 
That noise can accelerate phase space diffusion has been long known 
to dynamicists studying various maps, dating back at least to the work of 
Lieberman and Lichtenberg [5] in the 1970's. However, the details have not 
received all that much attention, especially for continuous systems.

But why should one care? Why is this phenomenon important in the real world?
The crucial observation here is that there is no such thing as a truly 
isolated system. Every system in nature is coupled to at least some degree to 
its surrounding environment. The important point then is that, in many cases,
one should expect that the coupling of a system to an external environment can 
be modeled as resulting in friction and noise, related by a 
Fluctuation-Dissipation Theorem [6,7] (although there {\it are} examples 
where such a picture {\it not} justified [8]). Indeed, the assumption of a 
system coupled by a Fluctuation-Dissipation Theorem to an environment, 
idealised as a heat bath characterised by some temperature ${\Theta}=k_{B}T$, 
is one powerful starting point for modern theories of nonequilibrium 
statistical mechanics (see, e.g., the textbook by Kubo, Toda, and 
Hashitsume [9]).

Most modeling of this sort involves the assumption of a composite entity of
system plus environment which is characterised by a time-independent 
Hamiltonian $H$. One might therefore worry that this picture does not
extend naturally to a cosmological setting where, in the average comoving 
frame, the Hamiltonian $H$ typically acquires an explicit time-dependence.
Fortunately, however, the assumption of a time-independent $H$ is not 
essential [10]. Following Caldeira and Legett [6,7], it is natural to 
write the composite Hamiltonian $H$ as a sum
\begin{equation}
H=H_{sys}+H_{bath}+H_{int},
\end{equation}
where the system Hamiltonian $H_{sys}$ is completely arbitrary, $H_{bath}$
is idealised as the Hamiltonian for a collection of linearised excitations,
i.e., ``phonons,'' with (in general) time-dependent frequencies, and the 
interaction $H_{int}$ is an arbitrary function of the system variables but 
linear in the bath variables. (These restrictions on $H_{bath}$ and $H_{int}$ 
assure that each mode of the environment is only weakly coupled to the system, 
so that the environment can be visualised as a thermal ``bath.'') In this 
setting, one can always integrate out the explicit dependence on bath 
variables to derive an exact nonlocal (in time) Langevin equation for the 
system; and, if $H_{bath}$ is time-independent, this exact equation (and, 
presumably, any reasonable Markov approximation thereunto) will satisfy a 
Fluctuation-Dissipation Theorem, regardless of the possible time-dependence 
of $H_{sys}$ and $H_{int}$. This implies in particular that, for the case
of a conformally static Friedmann cosmology, one has a cosmological 
Fluctuation-Dissipation Theorem whenever the environment can be approximated 
as a conformally coupled field, e.g., electromagnetic blackbody 
radiation.\footnote{
In this connection, it should be noted [11] that, in the dipole
approximation, the Hamiltonian describing the interaction of an 
electron with a radiation field is, when formulated in an inertial
frame, equivalent to the independent oscillator model [12,13],
which is perhaps the simplest ``realistic'' example of the Hamiltonian
(1). Transforming to comoving coordinates makes $H_{sys}$ and $H_{int}$
time-dependent, but the conformal invariance of the electromagnetic field
implies that $H_{bath}$ remains time-independent.}

The natural inference is that, in a variety of different settings, including
many relevant to astronomical systems, weak couplings to an external 
environment will result in non-Hamiltonian perturbations which could have 
important physical implications. However, the proper form for the noise is not 
always obvious. The simplest model in terms of which to couple a system to its 
surroundings, the independent oscillator model [12,13] (which can be shown 
to be equivalent to many other phenomenological models that have been 
considered in the past [11]), leads immediately to additive white noise, 
i.e. state-independent noise that is delta-correlated in time. However, this 
simple form for the noise is a direct consequence of the assumptions (i) that 
$H_{int}$ is linear in the system variables and (ii) that the bath phonons are 
characterised by an ohmic distribution, i.e., a spectral distribution 
${\propto}{\;}{\omega}^{2}d{\omega}$ with appropriate cutoffs. Allowing 
$H_{int}$ to involve the system variables in a more complicated fashion leads 
to multiplicative (i.e., state-dependent) noise; allowing for a different 
spectral density leads to coloured noise (i.e., noise that is not 
delta-correlated in time).

When considering a galaxy embedded in a rich cluster, one is confronted with
a system which, in many cases, is significantly impacted by its surrounding
environment. Particularly close encounters between galaxies in a cluster, e.g.,
those resulting in physical collisions, probably cannot be viewed as 
``random'' events. However, large numbers of relatively weak interactions 
probably {\it can} be viewed as a source of friction and noise, although there 
is no obvious reason why the noise associated with these interactions can be 
approximated as delta-correlated in time.

Another piece of physics which one might hope to model as friction and 
noise, again arising in galactic astronomy, is discreteness effects reflecting 
the fact that a galaxy is comprised of a collection of nearly point mass 
stars, rather than the smoothed-out continuum assumed in the context of a
description based on the collisionless Boltzmann (i.e., gravitational Vlasov)
equation. In the context of a smooth one-particle distribution function,
these discreteness effects are typically described by a Fokker-Planck, or 
Landau, equation, which involves a velocity-dependent coefficient of dynamical 
friction and multiplicative noise (diffusion) related by a self-consistent 
Fluctuation-Dissipation Theorem [14]. Superficially this source of friction 
and noise might seem completely different from the aforementioned effects 
associated with an external environment. However this is not really so! In 
this setting, one can view the full many-particle dynamics as the composite 
entity of system plus environment, the reduced one-particle dynamics as the 
system, and couplings to higher order correlations ignored in a collisionless 
description as interactions that serve as a source of friction and 
noise [15].

In the past, a good deal of work has focused on the effects of relatively
strong friction and noise in triggering barrier penetration and other phenomena
which proceed on the natural relaxation time $t_{R}$ associated with the
system's approach towards thermal equilibrium (see, e.g.,  [14,16] 
and numerous references cited therein.). This is {\it not} the problem
of interest here. Rather, the objective of the work described in this paper 
has been to focus on much weaker perturbations, where $t_{R}$ is much longer 
than any time scale of interest, and to determine the extent to which friction 
and noise have significant statistical effects on the evolution of ensembles 
of orbits already on time scales  ${\ll}{\;}t_{R}$. 

The numerical experiments described in the next two sections were performed
with the aim of answering three basic questions:
\begin{enumerate}
\item{} How should one visualise the effects of accelerated phase space
transport induced by friction and noise?
\item{} How does the size of the effect scale with the amplitude of the
perturbation?
\item{} To what extent do the details of the perturbation matter? One knows,
for example, that multiplicative noise can drive a system towards thermal 
equilibrium much more quickly than additive noise [17], and, as such, it 
would seem natural to ask whether multiplicative noise can also accelerate
diffusion through cantori and Arnold webs more than additive noise.
\end{enumerate}
\vskip .2in
\centerline{\bf INVARIANT AND NEAR-INVARIANT DISTRIBUTIONS}
\vskip .15in
The computations described in this paper involved integrating 
Langevin equations of the form
\begin{equation}
{d{\bf x}\over dt}={\bf v} \qquad {\rm and} \qquad
{d{\bf v}\over dt}=-{\nabla}{\Phi}-{\eta}{\bf v}+{\bf F},
\end{equation}
these corresponding to motion in a time-independent Hamiltonian 
$H=v^{2}/2+{\Phi}({\bf r})$
which is perturbed by friction and noise. The quantity ${\eta}$ represents a 
coefficient of dynamical friction which, in general, can be a nontrivial 
function of both ${\bf r}$ and ${\bf v}$. The quantity ${\bf F}$ is a 
``random'' force, idealised as Gaussian white noise, which is characterised 
completely by the statistical properties of its first two moments. 
Specifically,
\begin{equation} 
{\langle}F_{i}(t){\rangle}=0 \qquad {\rm and} \qquad
{\langle}F_{i}(t_{1})F_{j}(t_{2}){\rangle}=2{\eta}{\Theta}{\delta}_{ij}
{\delta}_{D}(t_{1}-t_{2}) ,
\end{equation}
where $i$ and $j$ label vector components and angular brackets denote an 
ensemble average. 
The first of these conditions ensures that the average force vanishes
identically. The second ensures that the autocorrelation function is
delta-correlated in both direction and time. The normalisation
in eq. (3) imposes a Fluctuation-Dissipation Theorem which ensures that,
for $t\to\infty$, an arbitrary ensemble of orbits evolved with eqs. (2) will 
approach a canonical distribution with temperature ${\Theta}$.

To date, integrations have focused on three specific two-dimensional 
potentials, namely
the sixth order truncation of the Toda lattice potential [18],
\begin{equation}
{\Phi}(x,y)= {1\over 2}{\Bigl(}x^{2}+y^{2}{\Bigr)}+x^{2}y-{1\over 3}y^{3} 
+{1\over 2}x^{4}+x^{2}y^{2}+{1\over 2}y^{4}+x^{4}y+{2\over
3}x^{2}y^{3}-{1\over 3}y^{5}+ $$ $${1\over 5}x^{6}+x^{4}y^{2}+{1\over
3}x^{2}y^{4}+{11\over 45}y^{6},
\end{equation}
\par\noindent
the so-called dihedral potential [19] for one particular set
of parameter values, i.e., 
\begin{equation}
{\Phi}(x,y)=-(x^{2}+y^{2})+{1\over 4}(x^{2}+y^{2})^{2}-{1\over 4}x^{2}y^{2},
\end{equation}
\par\noindent
and the sum of isotropic and anisotropic Plummer potentials [20]
for specified core radii and anisotropy parameters, i.e., 
\begin{equation}
V(x,y)=-{1\over {\Bigl(}c^{2}+x^{2}+y^{2}{\Bigr)}^{1/2}}
-{m\over {\Bigl(}c^{2}+x^{2}+ay^{2}{\Bigr)}^{1/2}},
\end{equation}
with $c=20^{2/3}{\;}{\approx}{\;}0.136$, $a=0.1$ and $m=0.3$.
In all three cases, the constants were so chosen that, in absolute units, a 
characteristic crossing time $t_{cr}{\;}{\sim}{\;}1$. This implies  
that, if one visualises these potentials as representing large galaxies like
the Milky Way, the Hubble time $t_{H}{\;}{\sim}{\;}100-200$.

As discussed more carefully elsewhere [20], these three potentials manifest 
very different symmetries. Indeed, the only obvious feature which they share 
is that, for a variety of energies, they admit significant measures of both 
regular and chaotic orbits, so that the chaotic phase space regions are 
significantly impacted by cantori. The fact that, nevertheless, similar 
qualitative results were obtained for orbits evolved in all three potentials 
can thus be interpreted as evidence that the basic conclusions are probably 
robust. 

Because all three potentials yielded similar results, the largest number of
calculations were performed for the dihedral potential, the most inexpensive
computationally, which corresponds physically to a slightly ``squared'' 
Mexican hat potential. In all cases, the orbits were computed using a fourth 
order Runge-Kutta algorithm, noise being implemented using an algorithm 
developed by Griner {\it et al} [21]. Most of the integrations were 
performed using a time step ${\delta}t=10^{-3}$. It was verified that a 
shorter time step ${\delta}t=10^{-4}$ does not yield significantly different 
results.

The first class of experiments to be performed involved tracking the
evolution of ensembles of chaotic initial conditions of fixed energy $E$, 
selected from some small phase space region in the center of the stochastic 
sea far from any important cantori. These initial conditions were
first evolved into the future in the absence of any friction or noise by 
integrating the deterministic Hamilton equations. They were then reintegrated
allowing for friction and noise of variable amplitude. All these experiments
assumed additive white noise and friction characterised by a constant ${\eta}$,
the two quantities being related by a Fluctuation-Dissipation Theorem. The
temperature was frozen at a value ${\Theta}{\;}{\sim}{\;}E$ and the 
amplitude of the perturbing influences was varied by systematically changing 
the value of ${\eta}$.

In the absence of friction and noise, such orbit ensembles exhibit a two-stage 
evolution [20,22]. The first stage involves a rapid coarse-grained evolution,
proceeding exponentially in time, towards a phase space distribution which is 
near-invariant in the sense that, once achieved, it only exhibits 
significant changes on a much longer time scale. Basically, this 
near-invariant distribution corresponds to a distribution characterised by a 
nearly constant number density in those portions of the constant energy 
phase space hypersurface that are not blocked by cantori 
and a near-zero density everywhere else. The second stage involves a much 
slower evolution towards what appears to be a true invariant distribution, as 
orbits in the ensemble diffuse through cantori to access phase space regions 
that were avoided systematically over shorter time scales. This final 
invariant distribution corresponds to a microcanonical population of the 
accessible chaotic regions, i.e., a uniform (in canonical coordinates) 
population of those portions of the phase space that are accessible to orbits 
with the specified initial conditions. 
The time scale for the first stage of the evolution is set at least 
approximately by the largest short time Lyapunov exponents for the orbits in 
the
ensemble, which determine how fast the ensemble will disperse. Typically this
time ${\sim}{\;}t_{cr}$. The time scale for the second stage is set by the 
time scale on which orbits diffuse through cantori, typically 
${\gg}{\;}t_{cr}$.

Suppose now that the orbits are perturbed by friction and noise with 
${\eta}{\;}{\sim}{\;}10^{-9}-10^{-4}$, the limiting values here corresponding, 
respectively, 
to the typical amplitude associated with discreteness effects in very large 
and very small galaxies [23,24,25]. In this case, one finds that the time
scale associated with the first stage of the evolution is essentially 
unchanged, i.e., ensembles still approach a near-invariant distribution on 
a time scale ${\sim}{\;}t_{cr}$, but that the time scale for the second stage
decreases dramatically! In the absence of friction and noise, the time scale
associated with diffusion through cantori typically satisfies
$t_{diff}({\eta}=0){\;}{\sim}{\;}10^{3}-10^{5}t_{cr}$, but even very weak 
friction and noise can decrease $t_{diff}({\eta})$ by orders of magnitude. 
For example, ${\eta}{\;}{\sim}{\;}10^{-9}-10^{-6}$ can result in a diffusion
time as short as ${\sim}{\;}100t_{cr}$, an interval which, for large galaxies,
corresponds to the Hubble time $t_{H}$. Indeed, for values of ${\eta}$ as 
large as ${\eta}{\;}{\sim}{\;}10^{-4}$, the diffusion time scale $t_{diff}$ 
is often so short that one cannot clearly distinguish between two different 
stages of evolution. For values of ${\eta}$ that large, noisy ensembles 
exhibit a rapid approach towards a near-invariant distribution that differs
significantly from the near-invariant distribution associated with a purely
Hamiltonian evolution but is comparatively similar to the true invariant
distribution associated with a Hamiltonian evolution.
Grey-scale plots comparing representative deterministic and noisy 
near-invariant 
distributions in the dihedral and truncated Toda potentials are exhibited, 
respectively, in FIGURES 8 in [25] and FIGURES 2 in [24].

In the absence of friction and noise, one anticipates that a generic ensemble
of initial conditions will ultimately evolve towards a microcanonical
distribution, i.e., a uniform population of the accessible portions of the
constant energy hypersurface, but this will only happen on the relatively long
time scale $t_{diff}({\eta}=0)$. Alternatively, if one allows for friction and 
noise and integrates for a time ${\sim}{\;}t_{R}$, one anticipates 
an evolution towards a canonical distribution with temperature ${\Theta}$.
Noisy integrations performed for a time ${\ll}{\;}t_{R}$ but still much longer 
than the time required to breech cantori will result in a near-invariant 
distribution that can be reasonably visualised as a slightly ``thickened'' 
version of a constant energy microcanonical distribution. Because $E$ is not 
exactly conserved, this near-invariant distribution is not exactly 
microcanonical, i.e., not proportional to a delta function in energy. However, 
because $E$ is almost conserved and the orbits have 
succeeded in breeching cantori, this noisy near-invariant distribution is much 
closer to the purely Hamiltonian invariant distribution than to either a 
canonical distribution or the purely Hamiltonian near-invariant 
distribution [25]. In this sense, weak friction and noise can accelerate
an approach towards a near-microcanonical equilibrium.

\vskip .2in
\centerline{\bf FIRST PASSAGE TIME EXPERIMENTS}
\vskip .15in
\par\noindent
To quantify the rate at which individual trajectories diffuse through 
cantori, a collection of first passage time experiments was also performed.
These involved four components:
\par\noindent
1. Select individual initial conditions corresponding in the absence
of friction and noise to {\it sticky} or {\it confined chaotic orbits}, i.e., 
chaotic orbits which, because of cantori, are trapped near regular regions 
for relatively long times. 
\par\noindent
2. Specify the form and amplitude of the friction and noise.
\par\noindent
3. For each choice of form and amplitude, perform a large number 
(${\sim}{\;}2000\,$-$\,5000$) of different noisy realisations of
the same initial condition; and, for each noisy realisation, determine the
time at which the orbit escapes through one or more cantori to become
unconfined.
\par\noindent
4. Analyse the data to extract $N(t)$, the fraction of
the orbits that have not yet escaped within a time $t$.

The results quoted below involve orbits in the dihedral potential (5) with
$E=10$ where, in the absence of any friction or noise, the diffusion time
$t_{diff}{\;}{\sim}{\;}1000$. Other choices of potential or energy can yield
results that differ quantitatively, but the principal qualitative conclusions
seem unchanged. Estimating when an orbit has escaped was done by identifying a 
``masked'' region in the configuration space and recording the first
time that the orbit left this region. That this mask criterion is 
reasonable was tested in two ways: (1) It was verified that changing 
slightly the shape and location of the mask had no appreciable effects. (2) 
For the case of purely Hamiltonian trajectories, escape from the masked
region was shown to correspond to an abrupt increase in the value of the
largest short time Lyapunov exponent. This is in accord with the fact that,
albeit still chaotic, confined chaotic orbits are less unstable exponentially
than are unconfined chaotic orbits [20].
One interesting variant of the preceding, also considered, involved tracking
a localised ensemble of initial conditions corresponding to confined chaotic 
orbits evolved into the future both with and without friction and noise. 
These experiments yielded results very similar to those obtained from multiple
integrations of individual initial conditions. 

Six different forms of friction/noise were considered, namely: 
(1) additive white noise and a constant coefficient of dynamical friction 
${\eta}$, related by a Fluctuation-Dissipation Theorem at temperature 
${\Theta}=E=10$;
(2) multiplicative white noise and dynamical friction with 
${\eta}={\eta}_{0}v^{2}$, related by a Fluctuation-Dissipation Theorem
with ${\Theta}=E=10$;
(3) multiplicative white noise and dynamical friction with 
${\eta}={\eta}_{0}v^{-2}$, again related by a Fluctuation-Dissipation Theorem
with ${\Theta}=E=10$; and 
(4) - (6) the same noises as in (1) - (3) but vanishing friction. In all
six cases, the individual noisy realisations were generated using the same
pseudo-random seeds.

In analysing the effects of friction and noise, attention focused on three
principal issues, namely:
\par\noindent
1. What is the functional form of $N(t)$, the fraction of the orbits that have
not yet escaped?
\par\noindent
2. How does $N(t)$ depend on the amplitude of the perturbation?
\par\noindent
3. To what extent does the form of the friction and the noise actually matter?

Overall, in these experiments escape is a two-stage process.
Early on, there are no escapes. All that one sees is that, as one might
expect [25,26], different noisy realisations of the same initial condition 
diverge exponentially at a rate set by the value of the largest short time 
Lyapunov exponent for the unperturbed deterministic trajectory. Eventually, 
however, once the noisy ensemble has dispersed to the extent that the root 
mean squared ${\delta}r_{rms}{\;}{\sim}{\;}1.0$, 
individual noisy orbits begin to escape through holes in the cantori. This 
onset of escape is a comparatively abrupt phenomenon, the interval during 
which the first 5\% of the orbits escape typically being only a small fraction 
of the time $T$ before the first escape occurs. It is also clear that, at least
early on, escapes can be well approximated as a Poisson process, with the
confined orbits becoming unconfined at a nearly constant rate, i.e., 
\begin{equation}
 N(t){\;}{\approx}{\;}\cases { N(0), & if $t{\;}{\le}{\;}T$; \cr
                      N(0)\,{\rm exp}\,[-{\Lambda}(t-T)], & if $t>T$.\cr}
\end{equation}

This behaviour is illustrated in FIGURES 1 (a) and (b), which exhibit 
${\rm ln}\,N(t)$ as a function of time $t$ for two different initial
conditions integrated in the presence of a constant ${\eta}$ and additive 
white noise. Each panel summarises multiple noisy realisations of a single
initial condition evolved for a total time $t=1024$ with ${\Theta}=10$.
The six different curves in each panel, each summarising $4000$ noisy
reasliations, represent six different values of ${\eta}$, namely 
${\rm ln}\,{\eta}=-9$, $-8$, $-7$, $-6$, $-5$, and $-4$.

It is clear from FIGURE 1 that, although ${\rm ln}\,N(t)$ originally decreases
linearly in time, it eventually develops nontrivial curvature indicating
that the escape rate is slowly {\it decreasing}. Exactly why this is so 
is not completely clear. However, two important points should be noted.
(1) In every case where it is observed, this curvature arises at a time
sufficiently late that changes in energy ${\delta}E$ have become appreciable,
${\sim}{\;}10\%$ or more. This suggests strongly that, at least in part, this 
change in escape rate reflects changes in the ``effective'' Hamiltonian phase 
space in which the noisy orbits evolve. 
(2) In at least some cases, the curvature reflects the fact that, because
of the perturbations, some originally chaotic orbits have become trapped by
{\it KAM} tori. 
(If, for the nonescapers, the friction and noise are turned off at some 
time ${\tau}$ and the trajectories integrated for a significantly longer
time, it becomes apparent that many of the orbits have become regular!).

So how, overall, does the form of $N(t)$ scale with ${\eta}$, the amplitude
of the perturbation? When probing the time $T$ required before escapes
begin or the rate ${\Lambda}$ at which escapes initially proceed after they
begin, one finds a roughly logarithmic dependence on ${\eta}$. In other words,
when plotting $T({\eta})$ or ${\Lambda}({\eta})$ the natural independent
variable, i.e., the abscissa, is ${\rm ln}\,{\eta}$, {\it not} ${\eta}$.
An example thereof is provided in FIGURES 2 a and b, which summarise
data generated from a single initial condition with friction and additive 
white noise related by a Fluctuation-Dissipation Theorem. The top panel 
exhibits $T(0.01)$, the time required for 1\% of the orbits in a $4000$ orbit 
ensemble to become unconfined. The lower panel exhibits the best fit value of 
the escape rate ${\Lambda}$ of eq. (7), as fit to the interval $T(0.01)<t<256$.
The curvature observed in both panels is statistically significant, so one
cannot assert that $T$ or ${\Lambda}$ are linear functions of 
${\rm ln}\,{\eta}$. However, it {\it is} clear that, overall, $T$ and 
${\Lambda}$ should be visualised as functions of ${\rm ln}\,{\eta}$ rather 
than ${\eta}$.

Perhaps the most important conclusion derived from these experiments is that
the computed $N(t)$ is nearly independent of the presence or absence of
friction, and that $N(t)$ is also largely independent of whether the noise is
additive or multiplicative! 

First perform $4000$ noisy realisations of the same initial condition, all 
with the same ${\Theta}$ and the same ${\eta}({\bf v})$, and analyse the
resulting data to extract $N(t)$. Then repeat these experiments with exactly
the same noise (generated from the same pseudo-random seeds!) but without
friction, and once again compute $N(t)$. A comparison of the two $N(t)$'s then
shows virtually no appreciable differences. At early times, there are 
absolutely
no statistically significant differences. Later on, one {\it can} see some
tiny differences. However, these can be attributed entirely to the fact that 
the energies of the orbits with and without friction will be slightly 
different, and that slightly different energies can give rise to slightly
different escape statistics. 

Comparing additive and multiplicative noise is a bit more subtle since one
must worry about normalisations. Suppose, however, that, when introducing 
multiplicative noise, one selects ${\eta}_{0}$ so that the ``average''
${\eta}{\;}{\equiv}{\;}{\eta}_{0}{\langle}v^{2}{\rangle}$ or
${\eta}{\;}{\equiv}{\;}{\eta}_{0}{\langle}v^{-2}{\rangle}$ coincides with the 
white noise constant ${\eta}$. In this case, one finds that the form of the
noise matters very little. Plots of $N(t)$ for additive white noise,
multiplicative noise ${\propto}{\;}v^{2}$, and multiplicative noise
${\propto}{\;}v^{-2}$ yield no statistically significant differences. 

Two examples of this behaviour are exhibited in FIGURES 3 (a) and (b) 
which compare the effects of additive and multiplicative
noise for two different initial conditions at two different perturbation 
levels. Each panel contains four curves, representing (1) additive white
noise and friction with a constant ${\eta}$, (2) the same additive white 
noise with vanishing friction, (3) multiplicative noise and friction with
${\eta}{\;}{\propto}{\;}v^{2}$, and (4) multiplicative noise and friction with
${\eta}{\;}{\propto}{\;}v^{-2}$. It is evident that, at late times, the curves 
do not completely overlap. However, it is also clear that none of the curves
is extremely different from the others.

These numerical experiments suggest two obvious inferences which, however,
remain to be checked more carefully for larger orbit ensembles, different
frictions and noise, and other potentials: (1) Smooth non-Hamiltonian 
perturbations like friction play only a minimal role in accelerating phase
space transport through cantori. (2) At least assuming that the noise is
white, its details seem comparatively unimportant. In particular, additive
noise and multiplicative noise depending on the orbital velocity ${\bf v}$
exhibit only minimal differences. Overall, when perturbing the Hamiltonian
trajectories what seems important is that the orbits be subjected to highly
``irregular'' perturbations that violate Liouville's Theorem at some given
amplitude. In this context, it should perhaps be noted explicitly that a
rapidly varying time-dependent Hamiltonian need not be as efficient in
triggering accelerated phase space transport as random noise. Specifically,
when considering a Hamiltonian of the form $H=H_{0}+{\epsilon}H_{1}(t)$,
with $H_{1}(t)$ periodic in time and ${\epsilon}$ an adjustable parameter, 
one finds in at least some cases [27] that, for periods ${\ll}{\;}t_{cr}$,
one must allow for relatively large values of ${\epsilon}$ to dramatically
accelerate diffusion through cantori.
\vskip .2in
\centerline{\bf WORK IN PROGRESS AND POTENTIAL IMPLICATIONS}
\vskip .15in
\par\noindent
The experiments described above still need to be generalised in two
important ways.

One obvious tack involves extending the computations to three-dimensional
systems. Arnold webs can serve as partial phase space obstructions in the
same sense as can cantori; and one might anticipate that friction and noise 
could accelerate phase space transport through such barriers in 
three-dimensional systems in the same ways as they do through cantori in 
two dimensions. This remains, however, to be checked. Indeed, for orbit 
ensembles in three-dimensional Hamiltonian systems, even the purely 
deterministic evolution towards an an invariant or near-invariant distribution 
is not completely understood. Preliminary investigations performed by Merritt 
and Valluri [28] suggest that the approach towards equilibrium can closely 
resemble what is observed in 
two-dimensional systems [20,22]. However, one would anticipate that, for
a generic three-dimensional system with two positive Lyapunov exponents, the
situation could be more complicated -- and interesting -- than in two
dimensions since unequal Lyapunov exponents could induce ``mixing'' that
proceeds in different directions at different rates [29]!

Another equally important objective is to allow for the effects of
coloured noise, where the autocorrelation function 
${\langle}F_{i}(t_{1})F_{j}(t_{2}){\rangle}$ is not delta-correlated in time. 
The assumption of singular, delta-correlated noise is an idealisation never 
exactly realised in nature, even when modeling high frequency phenomena like 
discreteness effects in systems interacting via short range forces; and the 
assumption seems especially unreasonable when trying to model objects like
galaxies embedded in a dense cluster environment where individual ``random''
interactions would seem characterised dimensionally by a time scale 
${\sim}{\;}t_{cr}$ or even larger! Allowing for coloured noise could also
be important by providing some insights into the question of exactly how and 
why non-Hamiltonian perturbations result in accelerated phase space transport.
Diffusion through cantori, either in the absence or presence of noise, must 
be related to some ``natural'' microscopic time scale(s), but the precise 
nature of these time scales has not yet been established. Systematically
increasing the autocorrelation time from zero (white noise) to values 
substantially larger will allow one to determine the point at which a
finite correlation time actually begins to matter.

In summary, the numerical experiments described in this paper lead to at
least three tentative conclusions:
\par\noindent
1. At least for systems that admit a coexistence of regular and chaotic 
behaviour, even weak couplings to an external environment, modeled as 
friction and noise, can dramatically accelerate evolution towards a 
(near-)microcanonical equilibrium. This suggest in particular that idealising 
a complex Hamiltonian system as a completely isolated entity may be a 
very bad idea.
\par\noindent
2. There is reason to think that the detailed form of the friction and 
noise is comparatively unimportant. When assessing the effects of friction
and noise in accelerating phase space transport, all that really matters 
may be the amplitude of the perturbation. If true, this would suggest that
it may not be all that hard to satisfactorily model the coupling of one's
system to its surrounding environment. The details which are hard to determine
may not be very important!
\par\noindent
3. ``Collisionality,'' i.e., discreteness effects, may be significantly
more important in galactic dynamics than generally recognised. For example,
such graininess could serve to destabilise quasi-equilibria which use confined 
chaotic orbits to support interesting structures such as bars [30] or 
triaxial cusps [31].
\vskip .2in
\centerline{\bf ACKNOWLEDGMENTS}
\vskip .15in
\par\noindent
It is a pleasure to acknowledge useful interactions with my collaborators,
Katja Lindenberg, Elaine Mahon, Ilya Pogorelov, and, especially, Salman Habib.
I am also grateful to James Meiss and Donald Lynden-Bell for useful comments
and critiques. The final draft of this manuscript was written at the Aspen
Center for Physics, the hospitality of which I acknowledge gratefully.
\vskip .2in
\vfill\eject
\centerline{\bf REFERENCES}
\vskip .1in
\par\noindent
1. AUBRY, S. \& G. ANDRE. 1978. 
{\it In} Solitons and Condensed Matter 
Physics. \par A. R. Bishop and T. Schneider, Eds.: 264. Springer, Berlin.
\par\noindent
2. MATHER, J. N. 1982.
Topology {\bf 21}: 457. 
\par\noindent
3. MACKAY, R. S., J. D. MEISS, \& I. C. PERCIVAL. 1984. 
Phys. Rev. Lett. {\bf 52}: 697.
\par\noindent
4. MACKAY, R. S., J. D. MEISS, \& I. C. PERCIVAL. 1984. 
Physica {\bf D 13}: 55.
\par\noindent
5. LIEBERMAN, M. A. \& A. J. LICHTENBERG. 1972. 
Phys. Rev. {\bf A 5}: 1852.
\par\noindent
6. CALDEIRA, A. O. \& A. J. LEGETT. 1983. Physica A {\bf 121}: 587. 
\par\noindent
7. CALDEIRA, A. O. \& A. J. LEGETT. 1983. 
Ann. Phys. (NY) {\bf 149}: 374.
\par\noindent
8. BARONE, P. M. V. B. \& A. O. CALDEIRA. 1991. 
Phys. Rev. {\bf A 43}: 57.
\par\noindent
9. KUBO, R., M. TODA, \& N. HASHITSUME. 1991. Statistical Physics II: 
Nonequilibrium \par Statistical Mechanics. Springer, Berlin, 2nd edition.
\par\noindent
10. HABIB, S. \& H. E. KANDRUP. 1992. 
Phys. Rev. {\bf D 46}: 5303.
\par\noindent
11. FORD, G. W., J. T. LEWIS, \& R. F. O'CONNELL. 1988. 
Phys. Rev. {\bf A 37}: 4419.
\par\noindent
12. FORD, G. W., M. KAC, \& P. MAZUR. 1965. 
J. Math. Phys. {\bf 6}: 504.
\par\noindent
13. ZWANZIG, R. 1973. 
J. Stat. Phys. {\bf 9}: 215.
\par\noindent
14. CHANDRASEHHAR, S. 1943. 
Rev. Mod. Phys. {\bf 15}: 1.
\par\noindent
15. KANDRUP, H. E. 1989.
Comments on Astrophys. {\bf 13}: 325.
\par\noindent
16. HONERKAMP, J. 1994. Stochastic Dynamics Systems. VCH Publishers, New York.
\par\noindent
17. LINDENBERG, K. \& V. SESHADRI. 1981. 
Physica {\bf A 109}: 481.
\par\noindent
18. TODA, M. 1967. 
J. Phys. Soc. Japan {\bf 22}: 431.
\par\noindent
19. ARMBRUSTER, D., J. GUCKENHEIMER, \& S. KIM. 1989. 
Phys. Lett. {\bf A 140}: 416. 
\par\noindent
20. MAHON, M. E., R. A. ABERNATHY, B. O. BRADLEY, \& H. E. KANDRUP. 1995.
\par
Mon. Not. R. Astr. Soc. {\bf 275}: 443.
\par\noindent
21. GRINER, A., W. STRITTMATTER, \& J. HONERKAMP. 1988. 
J. Stat. Phys. {\bf 51}: \par 95. 
\par\noindent
22. KANDRUP, H. E. \& M. E. MAHON. 1994. 
Phys. Rev. {\bf E 49}: 3735.
\par\noindent
23. KANDRUP, H. E. \& M. E. MAHON. 1995.
Ann. N. Y. Acad. Sci. {\bf 751}: 93.
\par\noindent
24. HABIB, S., H. E. KANDRUP, \& M. E. MAHON. 1996. 
Phys. Rev. {\bf E 53}: 5473. 
\par\noindent
25. HABIB, S., H. E. KANDRUP, \& M. E. MAHON. 1997. 
Astrophys. J. {\bf 480}: 155. 
\par\noindent
26. KANDRUP, H. E. \& D. E. WILLMES. 1994. 
Astron. Astrophys. {\bf 283}: 59.
\par\noindent
27. KANDRUP, H. E., R. A. ABERNATHY, \& B. O. BRADLEY. 1995. 
Phys. Rev. {\bf E \par 51}: 5287. 
\par\noindent
28. MERRITT, D. \& M. VALLURI. 1996. 
Astrophys. J. {\bf 471}: 82.
\par\noindent
29. KANDRUP, H. E. 1998. Mon. Not. R. Astr. Soc., submitted.
\par\noindent
30. WOZNIAK, H. 1993. {\it In} Ergodic Concepts in Stellar Dynamics. V. G.
Gurzadyan and \par D. Pfenniger, Ed. Springer, Berlin.
\par\noindent
31. MERRITT, D. \& T. FRIDMAN. 1996. 
Astrophys. J. {\bf 460}: 136.
\vfill\eject
\centerline{\bf FIGURE CAPTIONS}
\vskip .15in
\par\noindent
FIGS. 1 (a) $N(t)$, the fraction of confined chaotic orbits that have not yet
escaped to become unconfined, for ensembles of $4000$ noisy realisations of 
the same initial condition evolved in the dihedral potential with 
$E=10.0$, $x=0.0$, $y=1.3$, $v_{y}=1.75$, and $v_{x}=v_{x}(x,y,v_{y},E)>0$. 
Each orbit was subjected to additive white noise and friction 
with constant ${\eta}$, related by a Fluctuation-Dissipation Theorem 
with temperature ${\Theta}=10$. Passing from top to bottom at small $t$, the 
six curves represent ensembles with ${\eta}=10^{-9}$, $10^{-8}$, $10^{-7}$, 
$10^{-6}$, $10^{-5}$, and $10^{-4}$. (b) The same quantities generated for a 
different initial condition, namely $E=10.0$, $x=0.0$, $y=2.7$, and 
$v_{y}=2.25$. 
\vskip .1in
\par\noindent
FIGS. 2 (a) $T(0.01)$, the time required for 1\% of the members of an ensemble 
of $4000$ noisy realisations of the same unconfined chaotic orbit with 
$E=10.0$, $x=0.0$, $y=1.3$, $v_{y}=1.75$, and $v_{x}=v_{x}(x,y,v_{y},E)>0$ to 
become unconfined. Each orbit was subjected to additive 
white noise and friction with constant ${\eta}$, related by a 
Fluctuation-Dissipation Theorem with ${\Theta}=10$.
(b) ${\Lambda}$, the rate at which orbits in the ensemble escape, fit to the
interval $T(0.01)<t<256$.
\vskip .1in
\par\noindent
FIGS. 3 (a) $N(t)$, the fraction of confined chaotic orbits that have not yet
escaped to become unconfined, for ensembles of $4000$ noisy realisations of 
the same initial condition evolved in the dihedral potential with 
$E=10.0$, $x=0.0$, $y=1.1$, $v_{y}=3.35$, and $v_{x}=v_{x}(x,y,v_{y},E)>0$. 
Each orbit was evolved with ${\Theta}=10$ and ${\eta}_{0}=10^{-5}$. The four 
curves represent additive white noise and a constant ${\eta}$ (solid line),
additive noise but no friction (dashed), multiplicative noise and friction
with ${\eta}{\;}{\propto}{\;}v^{2}$ (dot-dashed), and multiplicative noise
and friction with ${\eta}{\;}{\propto}{\;}v^{2}$ (triple-dot-dashed). (b)
The same quantities generated for a different initial condition, namely
$E=10.0$, $x=0.0$, $y=1.3$, $v_{y}=1.75$, and $v_{x}=v_{x}(x,y,v_{y},E)>0$,
now allowing for ${\Theta}=10$ and ${\eta}_{0}=10^{-7}$.
\vfill\eject

\pagestyle{empty}
\begin{figure}[t]
\centering
\centerline{
        \epsfxsize=12cm
        \epsffile{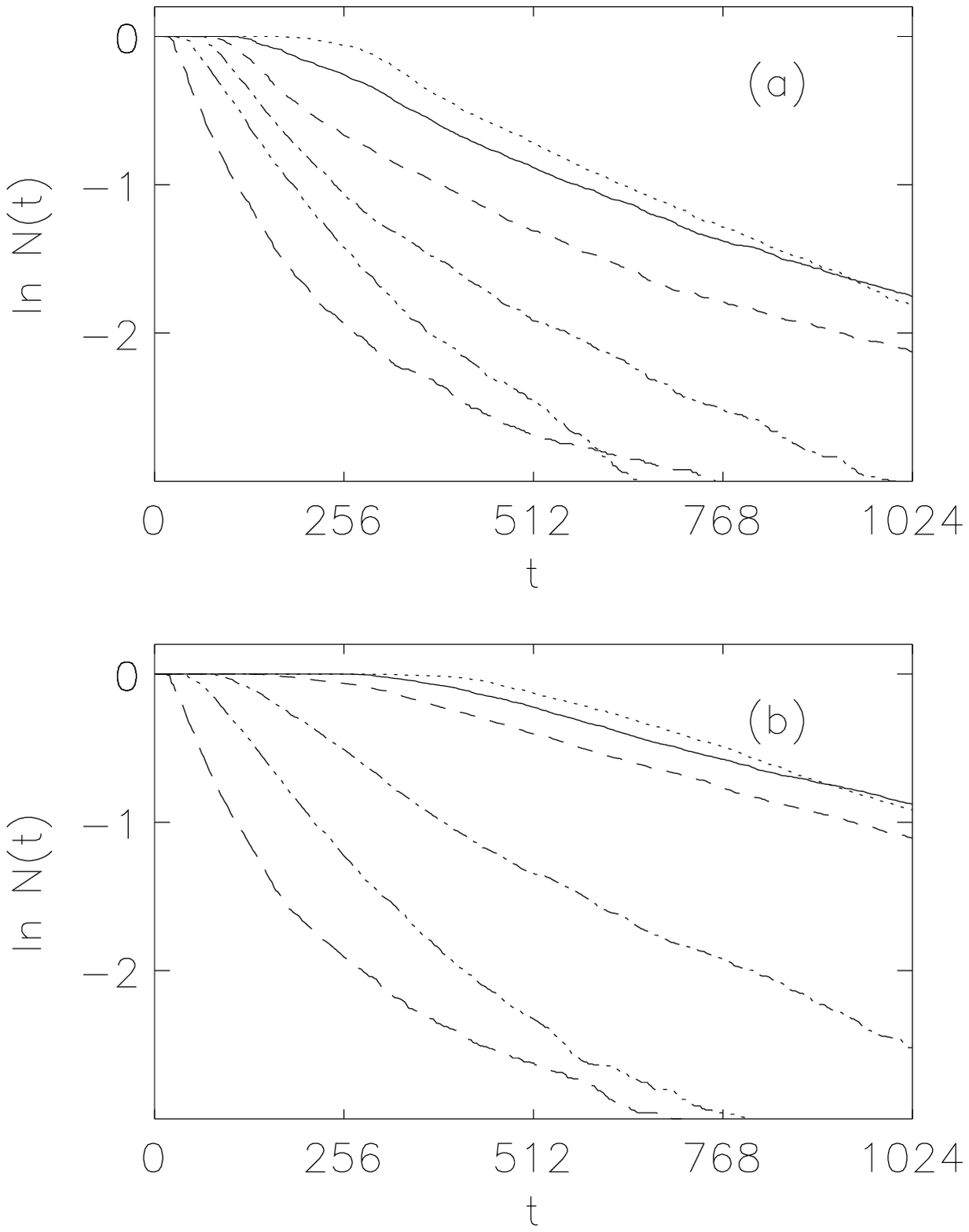}
           }
        \begin{minipage}{12cm}
        \end{minipage}
        \vskip -0.0in\hskip -0.0in
        \begin{center}\vskip .0in\hskip 0.5in
        Figure 1.
        \end{center}
\vspace{-0.2cm}
\end{figure}
\vfill\eject
\pagestyle{empty}
\begin{figure}[t]
\centering
\centerline{
        \epsfxsize=12cm
        \epsffile{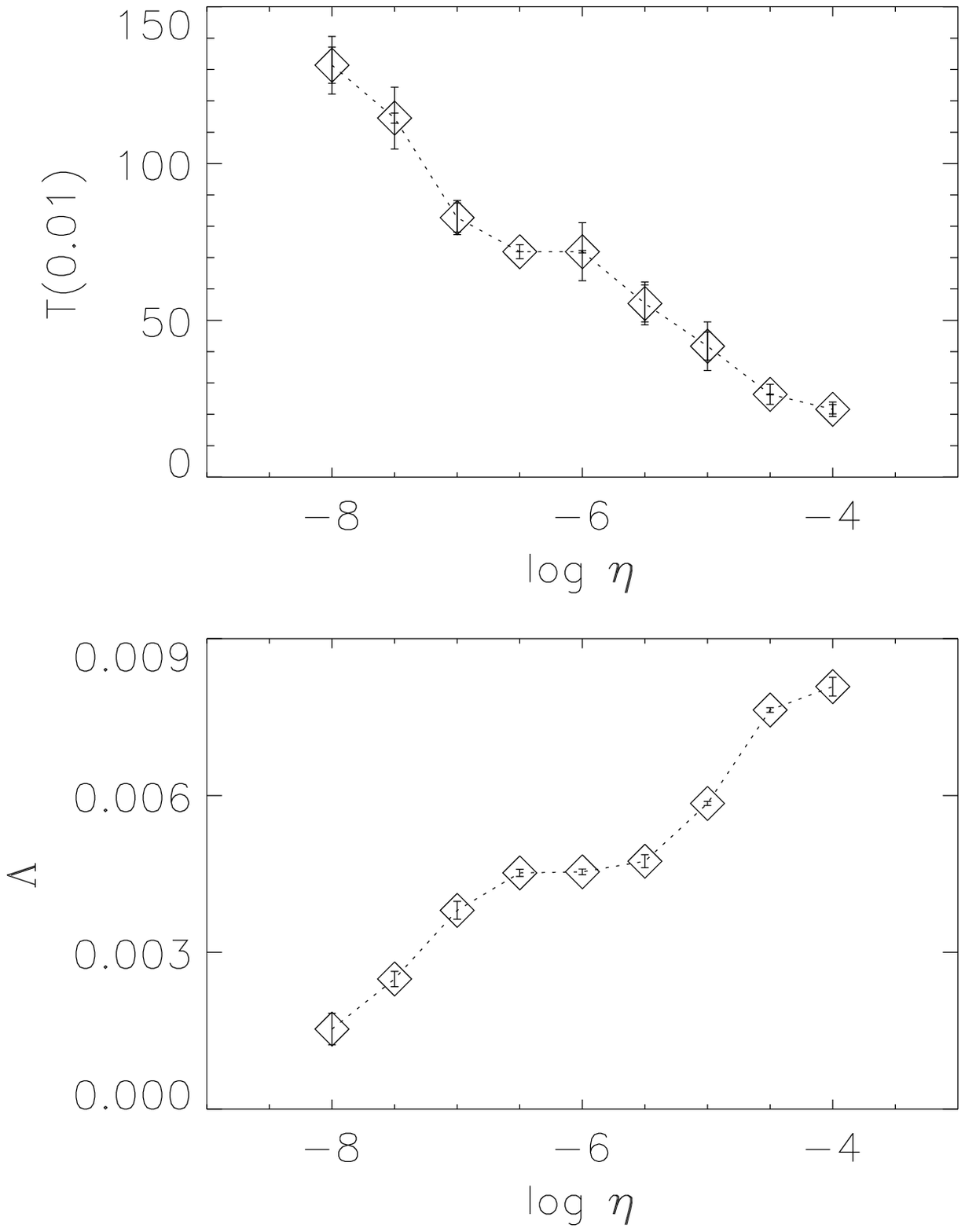}
           }
        \begin{minipage}{12cm}
        \end{minipage}
        \vskip -0.0in\hskip -0.0in
        \begin{center}\vskip .0in\hskip 0.5in
        Figure 2.
        \end{center}
\vspace{-0.2cm}
\end{figure}
\vfill\eject
\pagestyle{empty}
\begin{figure}[t]
\centering
\centerline{
        \epsfxsize=12cm
        \epsffile{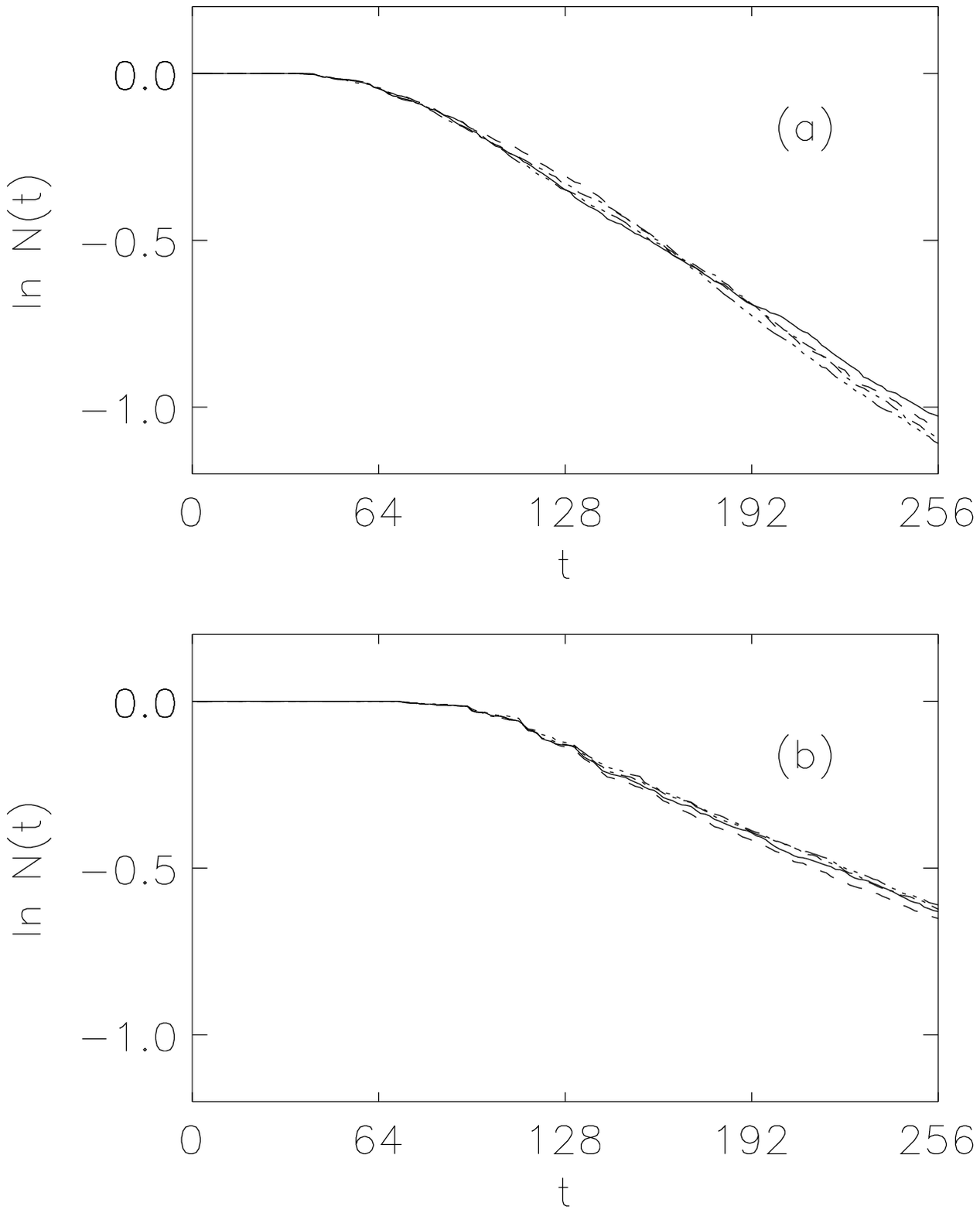}
           }
        \begin{minipage}{12cm}
        \end{minipage}
        \vskip -0.0in\hskip -0.0in
        \begin{center}\vskip .0in\hskip 0.5in
        Figure 3.
        \end{center}
\vspace{-0.2cm}
\end{figure}
\vfill\eject
\end{document}